 \documentclass[preprint ]{aastex} 

\shorttitle{Fe~XXII $I(11.92~{\rm\AA })/I(11.77~{\rm\AA })$ Density
Diagnostic}
\shortauthors{Mauche, Liedahl, \& Fournier}

 \slugcomment{Accepted for publication in 
             {\it Ap.J.\/} 2003 April 3}

\newcommand{\Mdot}{\dot{M}}
\newcommand{\Mwd}{M_{\rm wd}}
\newcommand{\Rwd}{R_{\rm wd}}

\newcommand{\lax}{{\lower0.75ex\hbox{ $<$ }\atop\raise0.5ex\hbox{ $\sim$ }}}
\newcommand{\gax}{{\lower0.75ex\hbox{ $>$ }\atop\raise0.5ex\hbox{ $\sim$ }}}
\newcommand{\pax}{{\lower0.75ex\hbox{ $\propto$ }\atop\raise0.5ex\hbox{ $\sim$ }}}

\begin{document}

\title{The Fe XXII I(11.92~\AA )/I(11.77~\AA ) Density Diagnostic Applied to
the Chandra High Energy Transmission Grating Spectrum of EX Hydrae}

\author{Christopher W.\ Mauche, Duane A.\ Liedahl,
    and Kevin B.\ Fournier}
\affil{Lawrence Livermore National Laboratory,
       L-473, 7000 East Avenue, Livermore, CA 94550; \\
       mauche@cygnus.llnl.gov, liedahl1@llnl.gov, fournier2@llnl.gov}


\begin{abstract}
Using the Livermore X-ray Spectral Synthesizer, which calculates spectral
models of highly charged ions based primarily on HULLAC atomic data, we
investigate the temperature, density, and photoexcitation dependence of the
$I(11.92~{\rm\AA })/I(11.77~{\rm\AA })$ line ratio of \ion{Fe}{22}. We find
that this line ratio has a critical density $n_{\rm c}\approx 5\times
10^{13}~\rm cm^{-3}$, is approximately 0.3 at low densities and 1.5 at high
densities, and is very insensitive to temperature and photoexcitation, so is a
useful density diagnostic for sources like magnetic cataclysmic variables in
which the plasma densities are high and the efficacy of the He-like ion density
diagnostic is compromised by the presence of a bright ultraviolet continuum.
Applying this diagnostic to the {\it Chandra\/} High Energy Transmission
Grating spectrum of the intermediate polar EX~Hya, we find that the electron
density of its $T_{\rm e} \approx 12$ MK plasma is $n_{\rm e}=1.0^{+2.0}_{-0.5}
\times 10^{14}~\rm cm^{-3}$, orders of magnitude greater than that typically
observed in the Sun or other late-type stars.
\end{abstract}

\keywords{atomic processes ---
          binaries: close ---
          stars: individual (EX Hydrae) ---
          stars: magnetic fields ---
          X-rays: binaries}


\section{Introduction}

Magnetic cataclysmic variables (mCVs; polars and intermediate polars) are
bright X-ray sources because the highly supersonic [free-fall velocity
$v_{\rm ff}=(2G\Mwd/\Rwd)^{1/2}\approx 3600~\rm km~s^{-1}$] material raining
down on the white dwarf primary passes through a strong shock where most of
its kinetic energy is converted into thermal energy and the plasma is heated to
a temperature $T_{\rm s} =3G\Mwd\mu m_{\rm H}/8k\Rwd\approx 200$~MK. Because of
the magnetic funneling of the mass lost by the secondary, the factor-of-four
density jump across the accretion shock, and the settling nature of the
post-shock flow (wherein the pressure is roughly constant and the
density scales inversely with the temperature; \citealt{fra92}), the hot
plasma in mCVs is expected to be dense. For a mass-accretion rate $\Mdot =
10^{15}~\rm g~s^{-1}$ ($L_{\rm X} =G\Mwd\Mdot/ 2\Rwd\approx 3\times
10^{31}~\rm erg~s^{-1}$) and a relative spot size $f=0.1$, the density of the
flow immediately behind the shock is $n= \Mdot/4\pi f\Rwd ^2\mu m_{\rm H}
(v_{\rm ff}/4)\approx 10^{13}~\rm cm^{-3}$.

The standard density diagnostic of high-temperature plasmas is the intensity
ratio $R\equiv f/i$ of the forbidden to intercombination lines of He-like ions
\citep{gab69, blu72, por01}, but this diagnostic is compromised in mCVs for two
reasons. First, the critical density of this ratio increases with $Z$ and hence
temperature---the opposite of the trend in the accretion column---so the $R$
line ratio is effective over only a very narrow slice of the accretion column.
Second, in mCVs and other ultraviolet-bright stars, photoexcitation competes
with collisional excitation to depopulate the upper level of the forbidden line,
so the $R$ line ratio can appear to be in the ``high-density limit'' if the
radiation field is sufficiently strong at the appropriate wavelengths: for a
plasma illuminated by a 30 kK blackbody, the $R$ line ratios of all elements
through Mg lie in the high-density limit, regardless of the density
\citep{mau02}.

To circumvent the density/photoexcitation ambiguity of the $R$ line ratio
of He-like ions, we undertook an investigation of potential density diagnostics
of lines of Fe L-shell ions that are observable in the high quality {\it
Chandra\/} High Energy Transmission Grating (HETG) spectrum of the intermediate
polar EX~Hya \citep{mau00, mau02}. Toward this end, \citet{mau01} investigated
the density, temperature, and photoexcitation dependence of the $I(17.10~{\rm
\AA })/I(17.05~{\rm\AA })$ line ratio of \ion{Fe}{17}, and found that the
anomalous ratio observed in EX~Hya can be explained if the electron density
of its $T_{\rm e}\approx 4$~MK plasma is $n_{\rm e}\gax 2\times 10^{14}~\rm
cm^{-3}$ or if the photoexcitation temperature $T_{\rm bb}\gax 55$~kK. In this
{\it Letter\/}, we investigate the density, temperature, and photoexcitation
dependence of the $I(11.92~{\rm\AA }) /I(11.77~{\rm\AA })$ line ratio of
\ion{Fe}{22}, which also is anomalous in EX~Hya. We find that this line ratio
is very insensitive to temperature and photoexcitation and that the observed
ratio can be explained if the electron density of EX~Hya's $T_{\rm e}\approx
12$ MK plasma is $n_{\rm e}\approx 1\times 10^{14}~\rm cm^{-3}$.

\section{Model Spectra}

The density sensitivity of the X-ray lines of \ion{Fe}{22} have been discussed
previously by \citet{dos73}, \citet{mas80}, \citet{dos81}, \citet{phi82},
\citet{faw87}, \citet{phi96}, and \citet{war98} in studies of the X-ray spectra
of solar and laboratory high-temperature plasmas. These diagnostics have not
received more attention because of the relative weakness of \ion{Fe}{22}
emission lines (the \ion{Fe}{22} ionization fraction peaks at just 22\% and
is greater than 1\% over the relatively narrow temperature range $T_{\rm e}
\approx 7$--26~MK; \citealt{maz98}), the relative weakness of $n\rightarrow
2$ lines for $n\ge 4$, and the fact that the critical densities of these
transitions are of order $10^{14}~\rm cm^{-3}$, larger than the densities
typically observed in the Sun, tokamaks, and electron beam ion traps. We
became interested in the density sensitivity of the $3\rightarrow 2$ lines of
\ion{Fe}{22} because the critical density of this transition is comparable to
the lower limit of the electron density derived from the \ion{Fe}{17}
$I(17.10~{\rm\AA })/ I(17.05~{\rm\AA })$ line ratio in the {\it Chandra\/}
HETG spectrum of EX~Hya, and because the \ion{Fe}{22} $3\rightarrow 2$
$I(11.92~{\rm\AA})/I(11.77~{\rm\AA })$ line ratio is anomalous (the 
11.92~\AA \ line is unusually strong) in the {\it Chandra\/} HETG spectrum of
EX~Hya compared to that of the bright late-type binaries HR 1099, Capella,
and $\sigma ^2$ CrB (\citealt{ayr01}, \citealt{phi01}, and \citealt{ost03},
respectively).

Before wading into the details, we refer the reader to Figure~1, which
illustrates the origin of the density dependence of the \ion{Fe}{22} spectrum.
At low densities ($n_{\rm e}\lax 10^{12}~\rm cm^{-3}$), the \ion{Fe}{22}
electron population is primarily in the $2s^22p_{1/2}$ ground state, and
collisional excitations are predominantly from the ground state into the
$2s2p^2$ manifold (8 levels total) and the $2s^23d_{3/2}$ level, both of which
decay primarily to ground, producing lines in the extreme ultraviolet (EUV)
($\lambda\lambda = 101$--217~\AA ) and at 11.77~\AA , respectively. However,
these levels also decay to the $2s^22p_{3/2}$ first-excited level with
approximately 15\% probability, producing lines in the EUV ($\lambda\lambda =
114$--349~\AA ) and at 11.93~\AA ; when {\it that\/} level decays to ground, it
produces a line in the ultraviolet (UV) ($\lambda = 846$~\AA ). As the density
increases ($n_{\rm e}\gax 10^{13}~\rm cm^{-3}$), electron population builds
up in the $2s^22p_{3/2}$ first-excited level because the M1 transition to
ground is slow ($A_{21}=1.47\times 10^4~\rm s^{-1}$). At high densities, the
$2s^22p_{3/2}$ first-excited level is fed primarily by radiative decays from
the $2s2p^2$ manifold. Collisional excitations out of the first-excited level
are primarily into the $2s^23d_{5/2}$ level, which decays primarily back to
the first-excited level, producing a line at 11.92~\AA . Consequently, the
\ion{Fe}{22} 11.92~\AA \ line is relatively strong in the X-ray spectra of
high-density plasmas.

To calculate quantitative models of the X-ray spectrum of \ion{Fe}{22}, we
employed the Livermore X-ray Spectral Synthesizer (LXSS), a suite of IDL
codes that calculates spectral models of highly charged ions based on Hebrew
University/Lawrence Livermore Atomic Code (HULLAC) atomic data. HULLAC
calculates atomic wavefunctions, level energies, radiative transition rates
$A_{\rm ji}$, and oscillator strengths $f_{\rm ij}$ according to the fully
relativistic, multiconfiguration, parametric potential method \citep{kla71,
kla77}. Electron impact excitation rate coefficients are computed
quasi-relativistically in the distorted wave approximation \citep{bar88}
assuming a Maxwellian velocity distribution. Our \ion{Fe}{22} model includes
electron impact excitation rate coefficients and radiative transition rates for
E1, E2, M1, and M2 decays for levels with principal quantum number $n\le 5$ and
azimuthal quantum number $l\le 4$ for a total of 228 levels. Using these data,
LXSS calculates the level populations for a given temperature and density
assuming collisional-radiative equilibrium; the line intensities are then
simply the product of the level populations and the radiative decay rates.

In a preliminary report, we \citep{mau03} used LXSS to calculate the X-ray
spectrum of \ion{Fe}{22} for electron densities $n_{\rm e}=10^{11}$ to
$10^{17}~\rm cm^{-3}$ and electron temperatures $T_{\rm e}=6.3$, 12.8, and
25.5 MK ($\approx {1\over 2}$, 1, and 2 times the temperature at which
the \ion{Fe}{22} ionization fraction peaks; \citealt{maz98}). At the
resolution of the {\it Chandra\/} HETG, the $2s^23d_{3/2}$--$2s^22p_{1/2}$
line at 11.77~\AA \ is blended with the ({\it very\/} much weaker)
$2s^23d_{5/2}$--$2s^22p_{1/2}$ line at 11.76~\AA , and the   
$2s^23d_{5/2}$--$2s^22p_{3/2}$ line at 11.92~\AA \ is blended with the (much
weaker) $2s^23d_{3/2}$--$2s^22p_{3/2}$ line at 11.93~\AA , so we report as
$I(11.92~{\rm\AA })/I(11.77~{\rm\AA })$ the ratio of the sum of these blends.
Figure 3 of \cite{mau03} shows that this line ratio has a critical density
$n_{\rm c}\approx 10^{14}~\rm cm^{-3}$, is approximately 0.3 at low densities
and 1.4 at high densities, and has only a weak temperature dependence.

\subsection{Modifications to LXSS: Collisional Excitation}

Motivated by the detailed study of \citet{fou01} of the density-sensitive Fe
lines in the EUV spectrum of the high-density Frascati Tokamak Upgrade
plasma, for this {\it Letter\/} we investigated the effect on the \ion{Fe}{22}
$I(11.92~{\rm\AA })/ I(11.77~{\rm\AA })$ line ratio of two modifications to
the collisional excitation data used in LXSS. First, we investigated the effect
of replacing, for all transitions between and among the $2s^22p$ and $2s2p^2$
levels of \ion{Fe}{22}, the electron impact excitation rate coefficients
computed with HULLAC with those of \citet{zha97}, computed with the relativistic
R-matrix method. At $T_{\rm e} =12.8$ MK, the Zhang \& Pradhan electron impact
excitation rate coefficients are larger than those of HULLAC by 1.95 for
the $2s^22p_{1/2}$--$2s^22p_{3/2}$ transition,  0.92--1.70 (average 1.27) for
the eight $2s^22p_{1/2}$--$2s2p^2$ transitions, 1.05--3.23 (average 1.55) for
the eight $2s^22p_{3/2}$--$2s2p^2$ transitions, and 0.45--8.99 (average 2.20)
for transitions between levels within the $2s2p^2$ manifold (but only
1.00--1.02 for the four $2s^22p$--$2s^23d$ transitions).  Despite the
significant differences between these excitation rate coefficients, we
find that for $T_{\rm e}=12.8$ MK the \ion{Fe}{22} $I(11.92~{\rm\AA })/
I(11.77~{\rm\AA })$ line ratio increases by 7\% at $n_{\rm e}=10^{14}~\rm
cm^{-3}$, but the low- and high-density limits are unaffected. Second, we
investigated the effect of adding proton excitations for transitions among the
levels of the $2s^22p$ and $2s2p^2$ configurations. This was accomplished by
fitting the proton impact excitation rate coefficients of \citet{fos97} to an
expression of the form $Q(T_{\rm e}) = 10^{-12}\, T_{\rm e}^{-1/2}
\sum_{i=1}^{3} A_i\,\exp(-B_i/T_{\rm e})$, where $Q(T_{\rm e})$ is the proton
impact excitation rate coefficient in units of $\rm cm^{3}~s^{-1}$, and we have
assumed that the proton and electron temperatures are equal. Adding proton
excitations to the LXSS population kinetics calculation under the assumption
that the proton number density $n_{\rm p} = 0.85\, n_{\rm e}$ (as is
appropriate for a high-temperature plasma with solar abundances), we find that
for $T_{\rm e}=12.8$ MK the \ion{Fe}{22} $I(11.92~{\rm\AA })/I(11.77~{\rm\AA 
})$ line ratio increases by 19\% at $n_{\rm e} =10^{14}~\rm cm^{-3}$, and the
high-density limit increases by 7\% to approximately 1.5.

With both of these modifications to LXSS, we recalculated the X-ray spectrum
of \ion{Fe}{22} for electron densities $n_{\rm e}=10^{11}$ to $10^{17}~\rm
cm^{-3}$ and electron temperatures $T_{\rm e}=6.3$, 12.8, and 25.5 MK. The
upper panel of Figure~2 shows that the model \ion{Fe}{22} $I(11.92~{\rm\AA })
/I(11.77~{\rm\AA })$ line ratio has a critical density $n_{\rm c}\approx 5
\times 10^{13}~\rm cm^{-3}$, is approximately 0.3 at low densities and 1.5 at
high densities, and has only a very weak temperature dependence.

\subsection{Modifications to LXSS: Photoexcitation}

To investigate the effects of photoexcitation on the \ion{Fe}{22} $I(11.92~{\rm
\AA })/I(11.77~{\rm\AA })$ line ratio, we added to the LXSS population
kinetics calculation the photoexcitation rates $(\pi e^2/m_{\rm e}c) f_{\rm ij}
F_\nu (T)$, where $F_\nu (T)$ is the spectrum of the photoexcitation radiation
field. For simplicity, we assume that $F_\nu (T) = (4\pi /h\nu ) B_\nu (T_{\rm
bb})$ (i.e., the radiation field is that of a blackbody of temperature $T_{\rm
bb}$) and the dilution factor of the radiation field is equal to $1\over 2$
(i.e., the X-ray emitting plasma is in close proximity to the source of the
photoexcitation continuum). With these assumptions, we used LXSS to calculate
the X-ray spectrum of \ion{Fe}{22} for an electron temperature $T_{\rm e}=
12.8$ MK, electron densities $n_{\rm e}= 10^{11}$ to $10^{17}~\rm cm^{-3}$,
and photoexcitation temperatures $T_{\rm bb}=20$, 40, 60, 80, and 100 kK. The
lower panel of Figure~2 shows that below $T_{\rm bb}=60$ kK the \ion{Fe}{22}
$I(11.92~{\rm\AA })/I(11.77~{\rm\AA })$ line ratio has essentially no
photoexcitation dependence, but as the photoexcitation temperature increases
the line ratio increases at low electron densities and the critical density
shifts to higher densities, while the high-density limit is unaffected.
Compared to the \ion{Fe}{17} $I(17.10~{\rm\AA })/I(17.05~{\rm\AA })$ line
ratio, the \ion{Fe}{22} $I(11.92~{\rm\AA })/I(11.77~{\rm\AA })$ line ratio
is insensitive to photoexcitation because the $2s^22p_{1/2}$--$2s^22p_{3/2}$
transition in the UV is not optically allowed ($f_{12}=3.15\times 10^{-6}$) and
the $2s^22p_{1/2}$--$2s2p^2$ transitions lie at such short wavelengths in the
EUV.

\section{Observations}

The {\it Chandra\/} HETG/ACIS-S observation of EX~Hya was performed between
2000 May 18 $\rm 9^h41^m$ and May 19 $\rm 2^h54^m$ UT for a total exposure of
59 ks. Extraction of the grating spectra and calculation of the effective
area files was accomplished with the CIAO 2.1 suite of software using the
reprocessed data products and new calibration data files (version R4CU5UPD8)
for sequence 300041. Various aspects of this spectrum have been discussed by
\citet{mau00, mau02} and \citet{mau01}; here we discuss only a tiny portion
($\lambda\lambda\approx 11.60$--12.05~\AA ) of the medium-energy grating (MEG)
spectrum as it bears on the relative strengths of the \ion{Fe}{22} $n = 3
\rightarrow 2$ emission lines. This spectrum is shown in Figure~3, where the
ordinate is in units of $\rm counts~s^{-1}~\AA ^{-1}$, the wavelength bin width
$\Delta\lambda =0.005$~\AA , and we have combined $\pm $ first order counts.
This portion of the spectrum contains the 11.77 and 11.92~\AA \ emission lines
of \ion{Fe}{22} as well as the 11.74~\AA \ emission line of \ion{Fe}{23}.

To determine the flux in these emission lines, we fit the MEG spectrum over
the $\lambda\lambda = 11.60$--12.05~\AA \ wavelength interval with a model
consisting of a linear background (to account for the thermal bremsstrahlung
continuum obvious in the broadband spectrum) and three Gaussians. Assuming a
common Gaussian width, the fitting function $f(\lambda ;\vec a)= a_1+
a_2(\lambda-\lambda_0) +\sum _{i=0}^{2} a_{5+2i}\exp [-(\lambda-a_{4+2i})^2
/2a_3^2]/\sqrt{2\pi }a_3$. To perform the fit, we separately accounted for
$\pm $ first orders and used $\Delta\lambda =0.005$~\AA \ bins, so there were
180 data points and 171 degrees of freedom (dof). The resulting fit, combining
$\pm $ first orders, is shown by the thick histogram in Figure~3. It gives
$\rm \chi^2/dof=104/171 =0.60$ and fit parameters as follows: Gaussian width
$a_3=1.4^{+1.6}_{-0.9}$ m\AA \ and line wavelengths and fluxes 
$I(11.740~{\rm\AA })=7.74\, (\pm 0.80)\times 10^{-5}$,
$I(11.776~{\rm\AA })=4.27\, (\pm 0.67)\times 10^{-5}$, and
$I(11.923~{\rm\AA })=4.52\, (\pm 0.68)\times 10^{-5}~\rm
photons~cm^{-2}~s^{-1}$, where the errors on the wavelengths are $\pm $ 1--2
m\AA \ and all errors are $1\,\sigma $ (68\% confidence) for one interesting
degree of freedom. Formally (e.g., \citealt{bev69}), the $I(11.92~{\rm\AA })/
I(11.78~{\rm\AA })$ line ratio equals $1.06\pm 0.23$ (68\% confidence), but
this propagation of errors ignores a subtle interplay between the error bars of
the numerator and denominator of this ratio. To generate more rigorous values
for the error bars of this ratio, we used a Monte Carlo technique with $10^6$
realizations to determine that the observed $I(11.92~{\rm\AA })/I(11.78~{\rm
\AA })$ line ratio equals 1.06 and the errors are
${+0.26}\atop{-0.20}$ (68\% confidence),
${+0.48}\atop{-0.32}$ (90\% confidence), and
${+0.89}\atop{-0.47}$ (99\% confidence). 
This best-fit line ratio and these
error envelopes are shown superposed on both panels of Figure~2.

\section{Analysis}

Figure~2 shows that the combination of our LXSS models and {\it Chandra\/}
HETG spectrum places significant constraints on the density of the $T_{\rm
e}\approx 12$ MK plasma in EX~Hya. Specifically, allowing for a factor-of-two
uncertainty in the electron temperature, the upper panel of Figure~2 shows
that the electron density
$n_{\rm e}=1.0^{+2.0}_{-0.5}\times 10^{14}~\rm cm^{-3}$ at 68\% confidence,
$n_{\rm e}\ge            3.2\times 10^{13}~\rm cm^{-3}$ at 90\% confidence, and
$n_{\rm e}\ge            1.7\times 10^{13}~\rm cm^{-3}$ at 99\% confidence.
Conversely, the lower panel of Figure~2 shows that the electron density can
be arbitrarily low if the photoexcitation temperature $T_{\rm bb}\gax 100$~kK,
but that alternative is excluded by UV observations of EX~Hya. First, assuming
that the distance to EX~Hya is 100 pc, that the radius of its white dwarf
is $\Rwd = 10^{9}$ cm, that its 1010~\AA \ flux density equals $2.5\times
10^{-13}~\rm erg~cm^{-2}~s^{-1}~\AA ^{-1}$ \citep{mau99}, and that 100\% of
this emission is due to a 100~kK hotspot (i.e., assuming that the white
dwarf, accretion disk, and accretion curtains do not radiate in the UV), the
fractional emitting area of the spot is approximately 0.5\%, far smaller
than is inferred from optical and UV light curves of mCVs. Second, fits to
EX~Hya's UV spectrum yield effective temperatures $T_{\rm eff}\approx 25$~kK
\citep{gre97, eis02}, far lower than that required to produce the anomalous
\ion{Fe}{22} $I(11.92~{\rm\AA })/I(11.77~{\rm\AA })$ line ratio via
photoexcitation.

\section{Discussion}

To date, four different spectroscopic diagnostics have provided evidence of
high densities in the X-ray--emitting plasma of the intermediate polar EX~Hya.
\citet{hur97} used the line ratio of the \ion{Fe}{20}/\ion{Fe}{23} 133~\AA \
blend to the \ion{Fe}{21} 129~\AA \ line observed in the 1994 {\it Extreme
Ultraviolet Explorer\/} spectrum of EX~Hya to infer $n_{\rm e}\gax 10^{13}~\rm
cm^{-3}$ under the assumption that the plasma temperature $T_{\rm e}=10$~MK.
Using the {\it Chandra\/} HETG spectrum of EX~Hya, Mauche (2002) showed 
that the He-like ion $R$ line ratios of O, Ne, Mg, Si, and S are all in their
high-density limit; \citet{mau01} used the \ion{Fe}{17} $I(17.10~{\rm\AA })
/I(17.05~{\rm\AA })$ line ratio to infer $n_{\rm e}\gax 2\times 10^{14}~\rm
cm^{-3}$ for its $T_{\rm e}\approx 4$~MK plasma; and we here used the
\ion{Fe}{22} $I(11.92~{\rm\AA })/I(11.77~{\rm\AA })$ line ratio to infer
$n_{\rm e}\approx 1\times 10^{14}~\rm cm^{-3}$ for its $T_{\rm e}\approx 12$
MK plasma. Of these diagnostics, the \ion{Fe}{22} line ratio is the most
reliable because it has the highest critical density and is the least
sensitive to temperature and photoexcitation. We conclude first that the
{\it Chandra\/} HETG spectrum of EX~Hya requires plasma densities that are
orders of magnitude greater than those typically observed in the Sun or
other late-type stars, and second that the  \ion{Fe}{22} $I(11.92~{\rm\AA })
/I(11.77~{\rm\AA })$ line ratio [like the \ion{Fe}{17} $I(17.10~{\rm\AA })
/I(17.05~{\rm\AA })$ line ratio] is a useful density diagnostic for sources
like mCVs in which the plasma densities are high and the efficacy of the
He-like ion density diagnostic is compromised by the presence of a bright UV
continuum. Finally, we note that our density determinations for EX~Hya are
consistent with those expected for plasma in the accretion column, where
$n_{\rm e}\pax  T_{\rm e}^{-1}$. Additional density diagnostics are needed to
begin to map the temperature/density profile of the accretion column of this
and other mCVs.

\acknowledgments

We thank H.~Tananbaum for the generous grant of Director's Discretionary Time
that made possible the {\it Chandra\/} observations of EX~Hya. Support for
this work was provided in part by NASA through {\it Chandra\/} Award Number
DD0-1004B issued by the {\it Chandra\/} X-Ray Observatory Center, which is
operated by the Smithsonian Astrophysical Observatory for and on behalf of NASA
under contract NAS8-39073. This work was performed under the auspices of the
U.S.~Department of Energy by University of California Lawrence Livermore
National Laboratory under contract No.~W-7405-Eng-48.



\begin{figure}
\figurenum{1}
\epsscale{0.53}
\plotone{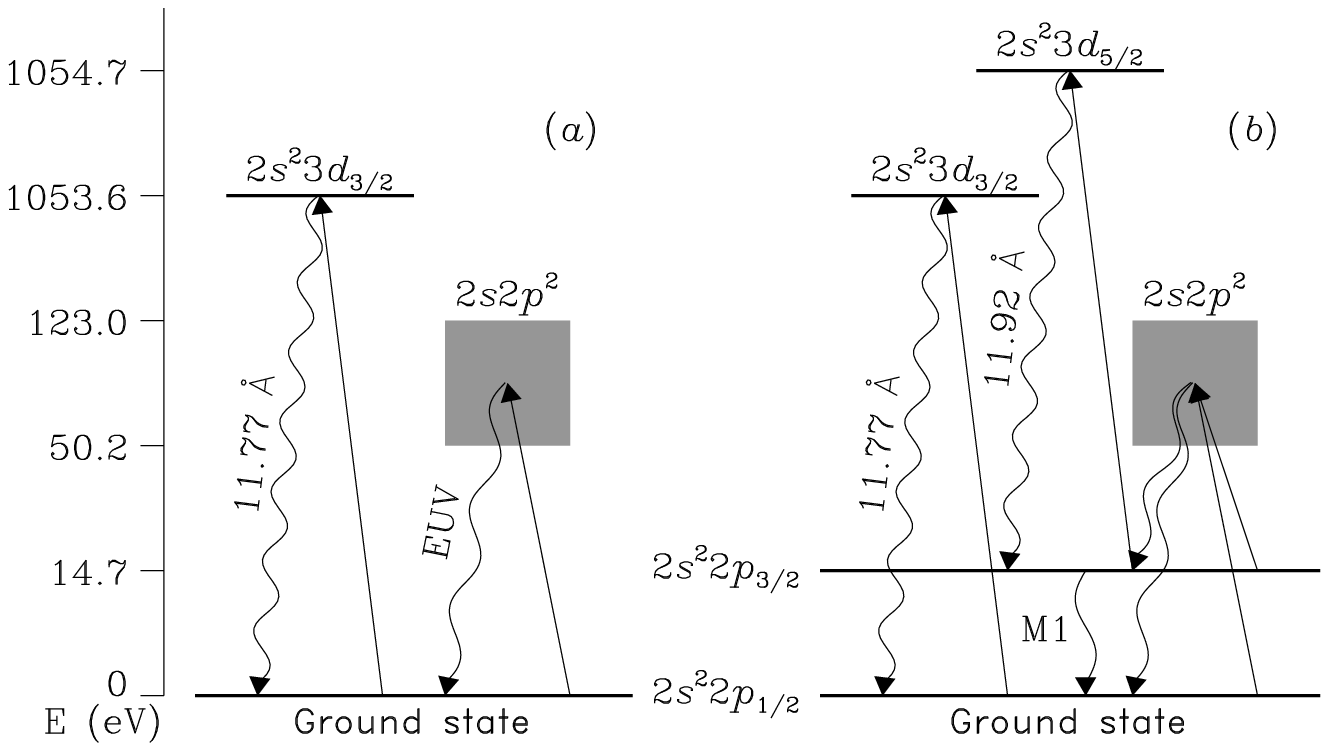}
\caption{
Schematic diagram of the dominant collisional and radiative processes of
\ion{Fe}{22} for electron densities $n_{\rm e}=10^{11}$ ({\it a\/}) and
$10^{17}~\rm cm^{-3}$ ({\it b\/}). Collisional excitations are indicated by
  upward-pointing straight lines; radiative decays are indicated by
downward-pointing wavy     lines and are labeled with the wavelength of the
transition in Angstroms.}
\end{figure}

\begin{figure}
\figurenum{2}
\epsscale{0.56}
\plotone{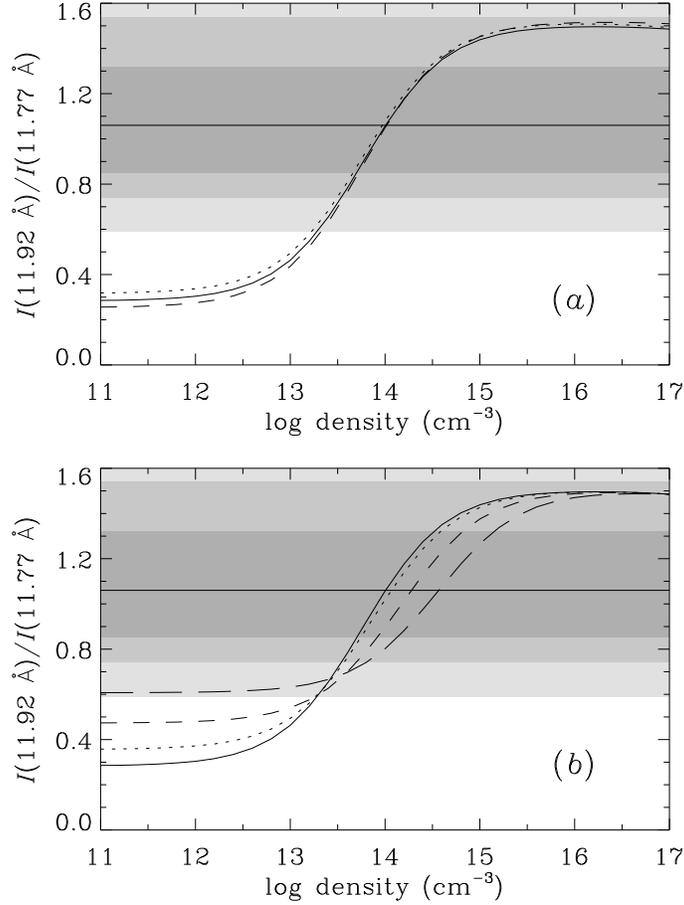}
\caption{
LXSS model \ion{Fe}{22} $I(11.92~{\rm\AA })/I(11.77~{\rm\AA})$ line ratio as
a function of electron density for ({\it a\/}) electron temperatures
$T_{\rm e}=6.3$, 12.8, and 25.5 MK ({\it dotted, solid, and dashed curves,
respectively\/}) and ({\it b\/}) electron temperature $T_{\rm e}=12.8$ MK and
photoexcitation temperatures $T_{\rm bb}=0$, 60, 80, and 100 kK ({\it solid,
dotted, short dashed, and long dashed curves, respectively\/}). Horizontal line
and the dark, medium, and light shaded stripes indicate respectively the value
and 68\%, 90\%, and 99\% confidence error envelopes of the line ratio measured
in EX~Hya.}
\end{figure}

\begin{figure}
\figurenum{3}
\epsscale{0.55}
\plotone{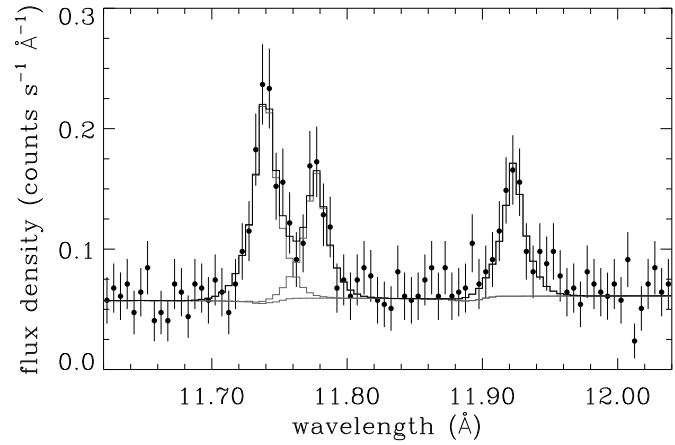}
\caption{
Detail of the {\it Chandra\/} MEG spectrum of EX~Hya in the neighborhood of
the \ion{Fe}{22} $3\rightarrow 2$ lines. Data are shown by the filled circles
with error bars and the net and separate contributions to the model fit are
shown by the histograms. Data combines $\pm $ first orders and are binned to
0.005~\AA .}
\end{figure}

\end{document}